\documentclass{jpp}
\usepackage{subeqn}
\usepackage{bm}
\usepackage{graphicx}
\title[Anisotropic structures in collisionless plasmas]
{Properties and evolution of anisotropic structures in collisionless plasmas}

\author[A.\,R. Karimov, M.\,Y. Yu and L. Stenflo]
{A.\ls R.\ns K\ls A\ls R\ls I\ls M\ls O\ls V\thanks{E-mail: alexanderkarimov999@gmail.com}$^1$, M.\ls Y.\ns Y\ls
U$^2$ and L.\ns S\ls T\ls E\ls
N\ls F\ls L\ls O$^3$}

\affiliation{$^1$Department of Electrophysical Facilities, National Research Nuclear University MEPhI, Kashirskoye shosse 31, Moscow, 115409, Russia and Institute for High Temperatures, Russian Academy
of Sciences, Izhorskaya 13/19, Moscow 127412, Russia \\[\affilskip] %%Institute for Fusion Theory and Simulation,
$^2$Department of Physics, Zhejiang University, 310027 Hangzhou,
China and Institut f\"ur Theoretische Physik I, Ruhr-Universit\"at
Bochum, D-44780 Bochum, Germany\\[\affilskip]
$^3$Department of Physics,
Link\"oping University, SE-58183 Link\"oping, Sweden}

\begin{document}
\newcounter{graf}
\maketitle

\begin{abstract}
A new class of exact electrostatic solutions of the Vlasov-Maxwell
equations based on the Jeans's theorem is proposed for studying
the evolution and properties of two-dimensional anisotropic
plasmas that are far from thermodynamic equilibrium. In
particular, the free expansion of a slab of electron-ion plasma
into vacuum is investigated.
\end{abstract}



%% 52.65.Ff  Fokker-Planck and Vlasov equation
%% 52.35.Mw  Nonlinear phenomena: waves, wave propagation, and other
% interactions (including parametric effects, mode coupling,
% ponderomotive effects, etc.)
%% 94.30.cs  Plasma motion; plasma convection
%% 95.30.Lz  Hydrodynamics

\section{Introduction}
There has been much interest in the properties of quasi-stationary
structures containing particles whose interaction is governed by
long-range, such as the gravitational or electrostatic, forces
[\cite{campa}-\cite{levin}]. Such structures are associated with
many phenomena, including solitons, shocks, vortices, nonlinear
waves, etc. in nature and the laboratory, and can often exist even
far from thermodynamic equilibrium because of absence collisional
or turbulent relaxation within the time scale of interest. In
particular, they can appear in highly rarefied plasmas not near
thermodynamic equilibrium. However, depending on its initial
distribution, a collisionless plasma can often still evolve into
quasi-stationary states because of the presence of the
self-consistent, or averaged, electrostatic field of the
individual charged particles [\cite{s04}-\cite{levin}]. However,
unlike collisional relaxation, which tends to drive the system
towards thermodynamic equilibrium, in collisionless relaxation any
initial energy imbalance among the different degrees of freedom
can be preserved, causing the system to evolve in a preferred
direction in the physical or phase space. Moreover, in plasmas the
motion of both the electrons and ions can play important roles in
the evolution, even though they are on very different time scales
because of the much larger ion mass. This is because the initial
or short-time behavior can often determine the pathway and thus
the asymptotic behavior of the highly nonlinear evolution (see,
e.g., Refs. [\cite{s04,ls05,es06}] and the references therein).
Complex behavior can also be expected for magnetized plasmas,
which are anisotropic and a large number of different modes of
collective motion can exist [\cite{clemmow69}].

In order to investigate the evolution and properties of
collisionless plasmas, we shall construct time-dependent non-Maxwellian
distribution functions satisfying the two-dimensional
Vlasov-Maxwell equations. In general, for initial states far from
equilibrium the convection terms in the governing equations are
not small [\cite{s04,kuz_96,bohr,kiessling}]. They in fact
determine the asymptotic behavior, which (if exists)
is usually still not near equilibrium (see, e.g.,
[\cite{taranov}-\cite{sch15}]). For such problems it is necessary
to use a fully nonlinear formulation. In this paper, we shall
invoke the Jeans's theorem [\cite{clemmow69}-\cite{pec12}] to
consider the two-dimensional (2D) evolution of a collisionless
electron-ion plasma slab, in particular, its expansion into vacuum.

\section{Formulation of the kinetic problem}\suppressfloats
We are interested in the non-relativistic electrostatic evolution
of a finite 2D unmagnetized electron-ion plasma slab. The
distribution functions $f_s=f_s(t,\bm{r},\bm{v})$, where $s=e,i$
for the electrons and ions, respectively, are governed by the
corresponding Vlasov and reduced Maxwell equations
\begin{equation} % ------------ (1_vl) -------------
\partial_t f_s + \bm{v}\cdot\nabla_{\bm{r}}f_s+{q_s\over
m_s}\bm{E}\cdot\nabla_{\bm{v}}f_s=0, \label{1_vl}
\end{equation}
\begin{equation} % ------------ (2_vl) -------------
\nabla\cdot\bm{E}=4\pi\sum_sq_sn_s, \label{2_vl}
\end{equation}
\begin{equation} % ------------ (f_vl) -------------
\nabla\times\bm{E}=0, \label{f_vl}
\end{equation}
\begin{equation} % ------------ (3_vl) -------------
{\partial_t\bm{E}}=-4\pi\sum_sq_s\bm{j}_s,
\label{3_vl}
\end{equation}
where $\bm{E}$ is the electrostatic field, and $n_s=\int
f_s(t,\bm{r},\bm{v})d\bm{v}$ the number density,
$\bm{j}_s=\int\bm{v}f_s(t,\bm{r},\bm{v})d\bm{v}$ the flux, and
$q_s$ the charge, of the $s$ particles. The reduced (no
displacement current and magnetic field perturbation) Maxwell
equations correspond to the Darwin approach for open-boundary
electrostatic problems in plasma physics [\cite{darwin}]. The
approach is particularly useful for considering electrostatic
phenomena in complex laboratory and space plasmas since by using
the reduced current equation, one can avoid solving the Poisson's
equation, which requires stringent boundary conditions
[\cite{arfken}]. In the Darwin approach, the electrostatic nature
of the problem is preserved by proper formulation of the initial
condition.

As mentioned, we shall consider the moving-boundary problem of the
expansion of a two-dimensional plasma slab far from thermal
equilibrium., i.e., we look for solutions of
(\ref{1_vl})--(\ref{3_vl}) in the form
\begin{equation} % ------------ (0_vl) -------------
f_{s}=f_{s}(t,x,y,v_x,v_y) > 0
\label{0_vl}
\end{equation}
defined in the region $\Gamma=\{(x,y), \mid x\mid \leq X_s(t),
\mid y\mid \leq Y_s(t)\}$, where $X_s(t)$ and $Y_s(t)$ denote the
average fronts, or boundaries (to be defined more precisely
later), separating the $s$ particles from the vacuum. The initial
slab can thus be defined by $X_s(t=0)=Y(t=0)=L$, where $L>0$ is
the initial dimensions of the plasma slab in the $x$ and $y$
directions. Accordingly, as initial condition we take
\[f_s(t=0,\bm{r},\bm{v}) =\left\{
\begin{array}{rl}
f_{0s}(v_x,v_y)>0, & \mid x\mid \leq L, \mid y\mid \leq L \\
0,            & \mid x\mid > L, \mid y\mid > L,
\end{array}
\right.\]
where the initial distribution function $f_{0s}(v_x, v_y)$ has finite moments
$$\left|\int_{-\infty}^{+\infty}\bm{v}^kf_{0s}(\bm{r},\bm{v})d\bm{v}\right|< \infty,
\quad \quad k=0,1,2,\ldots\/.$$
The plasma is assumed to be initially neutral and at rest, so that we have
\begin{equation} % ------------ (ad_couple) -------------
\int_{-\infty}^{+\infty}f_{0s}(\bm{r},\bm{v})d\bm{v}=1 \quad {\rm
and} \quad \int_{-\infty}^{+\infty}\bm{v}f_{0s}(\bm{r},
\bm{v})d\bm{v}=0\/. \label{ad_couple}
\end{equation}

It is convenient to normalize (\ref{1_vl})--(\ref{3_vl}) by
$$\bar{t}= \omega_{pe}t, \quad \bar{\bm{r}}={\bm{r} \over L},
\quad \bar{\bm{v}}={\bm{v} \over v_0}, \quad \bar{\bm{E}}={\bm{E}
\over E_0}, \quad \bar{n}_s= {n_s \over n_0},$$ where
$\omega_{pe}=\sqrt{4\pi n_0e^2/m_{e}}$ is the electron plasma
frequency, $E_0=4\pi en_0L$, $v_0=L\omega_{pe}$, $n_0$ is the
initial plasma density ($n_0=n_e=n_i$), and $-e$ and $m_e$ are the
charge and mass of the electron, respectively. For clarity, in the
following we shall omit the overhead bars. The normalized
equations are then
\begin{equation} % ------------ (1_vl) -------------
\partial_t f_s+\bm{v}\cdot\nabla_{\bm{r}}f_s+{Q_s\over
M_s}\bm{E}\cdot\nabla_{\bm{v}}f_s=0, \label{1_vln}
\end{equation}
\begin{equation} % ------------ (2_vl) -------------
\nabla\cdot\bm{E}=\sum_s Q_s n_s \label{2_vln}
\end{equation}
\begin{equation} % ------------ (3_vln) -------------
{\partial_t\bm{E}}=-\sum_s Q_s\bm{j}_s,
\label{3_vln}
\end{equation}
where $Q_e=-1$, $M_e=1$ and $Q_i=1$, $M_i=m_i/m_e$, and $m_i$ is the ion mass.

The plasma is assumed to be symmetric with respect to $(x=0,
y=0)$. It is therefore sufficient to consider only the upper half
of $\Gamma$. Accordingly, we can write
\begin{equation} % ------------ (incon_vln) -------------
f_{s}(t=0,x,y,v_x,v_y)= f_{0s}(a^0_{xs}v_x+b^0_{xs}v_y, a^0_{ys}v_x+ b^0_{ys}v_y),
\label{incon_vln}
\end{equation}
where $a^0_{xs}, b^0_{xs}, a^0_{ys}$, and $b^0_{ys}$ are
constants, defined in the initial region $\Gamma(t=0)=\{(x,y),
0\leq x\leq 1, 0\leq y\leq 1\}$. We see that the system is not in
thermodynamic (Maxwellian) equilibrium and there is a preferred
direction for the evolution, given by the constant coefficients
$a^0_{xs}, b^0_{xs}, a^0_{ys}$, and $b^0_{ys}$. The latter are
determined by the self-consistent electrostatic field as well as
external fields, if any.

\section{The invariants}
According to the Jeans's theorem, solutions of the Vlasov-Maxwell equation can be written as
\begin{equation} % ------------ (1_vlj) -------------
f_s=f_s(I_{1 s}, I_{2 s}, \ldots, I_{K s})\label{1_vlj}
\end{equation}
where $I_{1 s}, I_{2 s}, \ldots, I_{K s}$ are the invariants of
motion, i.e., they remain constant along the trajectory of a
particle, or along a phase-space characteristic of Eq.
(\ref{1_vln}), even though they can be functions of time, space,
and velocity.

We start from a simple case, where the invariants $I_{l s}$ are
linear function of ${\bf v}$, namely
\begin{equation} % ------------ (7_vl) -------------
I_{l s} = a_{l s}(t)v_{x}+b_{ls}(t)v_{y}+c_{ls}(t)x+d_{ls}(t)y + h_{ls}(t),
\label{7_vl}
\end{equation}
where the coefficients $a_{ls}(t)$, $b_{ls}(t)$, $c_{ls}(t)$,
$d_{ls}(t)$, and $h_{ls}(t)$ depend only on time. The Ansatz
(\ref{7_vl}) corresponds to presetting the spatial structures of
the self-consistent and the external fields (if any), and thereby
also the particle densities, currents, etc. These parameters have
to be obtained by trial and error, such that $d_tI_{ls}=0$ along
the particle trajectory [\cite{ll,struc}]. That is, the
time-dependent coefficients should exist and satisfy
(\ref{1_vln})-(\ref{3_vln}).

The equations for the coefficients $a_{ls}(t)$ to $h_{ls}(t)$ can
be obtained by substituting  (\ref{1_vlj}) into (\ref{1_vln}):
\begin{equation}\label{xx}
\sum_l \left[G_s(\bm{x},\bm{v})+\frac{Q _s}{M_s}\left(a_{ls}E_x
+b_{ls}E_y\right)\right]\partial_{I_{ls}}f_s=0,
\end{equation}
where
$$G_s(\bm{x},\bm{v})=\dot{a}_{ls}v_{x}+\dot{b}_{ls}v_{y}
+\dot{c}_{ls}x+\dot{d}_{ls}y+c_{ls}v_x+d_{ls}v_y+\dot{h}_{ls},$$
and the overhead dot denotes the time derivative. Since
$\partial_{I_{ls}}f_s$ should be independent for $s=e,i$, Eq.\
(\ref{xx}) is satisfied if
\begin{equation} % ------------ (8_vl) -------------
G_s(\bm{x},\bm{v})+\frac{Q_s}{M_s}\left(a_{ls}E_x+b_{ls}E_y\right)=0.
\label{8_vl}
\end{equation}
% is valid for $l=x,y$.
Since the space and velocity coordinates are independent in the
phase space, we obtain
%% \begin{equation} % ------------ (9_vl) -------------
$c_{ls}=-\dot{a}_{ls}$, $d_{ls}=-\dot{b}_{ls}$,
%%\label{9_vl}
%%\end{equation}
and
\begin{equation} % ------------ (10_vl) -------------
\ddot{a}_{ls}x+\ddot{b}_{ls}y-\dot{h}_{ls}=\frac{Q _s}{M_s}\left(a_{ls}
E_x+b_{ls}E_y\right). \label{10_vl}
\end{equation}

The initial conditions (\ref{incon_vln})
\begin{equation} % ------------ (2_vlj) -------------
a_{ls}(t=0)= a^0_{ls}, \quad b_{ls}(t=0)= b^0_{ls},
\quad h_{ls}(t=0)=\dot{a}_{ls}(t=0)= \dot{b }_{ls}(t=0)=0\/.
\label{2_vlj}
\end{equation}
allow us to define the number $K$ (here not more than four) of
constants motion. In fact, the set (\ref{7_vl}) is a system of
linear algebraic equations relating $I_{l s}$ to $v_x$ and $v_y$.
Accordingly, the four invariants $I_{l s} \neq 0$ at any time
uniquely determine $v_x$ and $v_y$. However, from (\ref{2_vlj}) we
have $d_s(t=0)=c_s(t=0)= h_{ls}(t=0)=0$ at $t=0$, so that the rank
of the matrix with $a_{l s}, b_{ls}, c_{ls}, d_{ls}$, and $h_{ls}$
is not more than 2. That is, there are only two independent
equations. The distribution functions can then be rewritten as
functions of the invariants
\begin{equation} % ------------ (6_vl) -------------
f_s(t,\bm{r},\bm{v})=f_s(I_{xs},I_{ys}), \label{6_vl}
\end{equation}
where $l=x,y$.

The case $K=1$ is singular: $I_{xs}$ and $I_{ys}$ are not linearly independent. Nevertheless,
it is still realistic and shall thus be separately considered later.

\section{The case $K=2$}
We first investigate the case $K=2$, and use (\ref{6_vl}) to determine
$a_{ls}$, $b_{ls}$, and ${\bf E}$. The particle densities and
fluxes can be expressed as integrals in $I_{xs}$ and $I_{ys}$ (in
place of $v_x$ and $v_y$) (see Appendix \ref{AppendixB})
\begin{equation} % ------------ (11_vl) -------------
n_s=\frac{1}{\Lambda_s}, \label{11_vl}
\end{equation}
which verifies the Ansatz that the densities $n_s$ are functions of time
only. Moreover, we have
\begin{equation} % ------------ (12_vl) -------------
j_{xs}=\frac{\dot{a}_{xs}b_{ys}-b_{xs}\dot{a}_{ys}}{\Lambda_s^2}x+
\frac{\dot{b}_{xs}b_{ys}-b_{xs}\dot{b}_{ys}}{\Lambda_s^2}y
+\frac{b_{xs}h_{ys}-b_{ys}h_{xs}}{\Lambda_s^2}, \label{12_vl}
\end{equation}
\begin{equation} % ------------ (13_vl) -------------
j_{ys}=\frac{\dot{a}_{ys}a_{xs}-a_{ys}\dot{a}_{xs}}{\Lambda_s^2}x+
\frac{a_{xs}\dot{b}_{ys}-a_{ys}\dot{b}_{xs}}{\Lambda_s^2}y
+\frac{a_{ys}h_{xs}-a_{xs}h_{ys}}{\Lambda_s^2}, \label{13_vl}
\end{equation}
where $\Lambda_s=a_{xs}b_{ys}-a_{ys}b_{xs}$.

From (\ref{2_vln}) and (\ref{11_vl}), we obtain
\begin{equation} % ------------ (16_vl) -------------
{\partial_x E_x}+{\partial_y E_y}=\sum_s{q_s\Lambda_s^{-1}},
\label{16_vl}
\end{equation}
where the right-hand side is only a function of $t$. One can then
write
\begin{equation} % ------------ (17_vl) -------------
E_x=A(t)x+B(t)y+H(t), \quad  E_y=C(t)x+D(t)y+F(t),
\label{17_vl}
\end{equation}
where the time-dependent functions $A$, $B$, $C$, $D$, $H$, and
$F$ still have to be determined. Substituting (\ref{17_vl}) into
Eq. (\ref{f_vl}) we get
\begin{equation} % ------------ (cb_vl) -------------
C(t)=B(t). \label{cb_vl}
\end{equation}
Eq. (\ref{10_vl}) then becomes
\begin{equation} % ------------ (18_vl) -------------
\ddot{a}_{ls}x+\ddot{b}_{ls}y-\dot{h}_{ls}
=\frac{Q_s}{M_s}\left[(a_{ls}A+ b_{ls} B)x+(a_{ls}B+b_{ls}D)y
+(a_{ls}H+b_{ls}F)\right]. \label{18_vl}
\end{equation}
The terms involving the space coordinates $x$ and $y$ can now be
separated. Accordingly, we have
\begin{equation} % ------------ (29_vl) -------------
\ddot{a}_{ls}=\frac{Q_s}{M_s} (a_{ls}A+b_{ls}B), \quad \ddot{b}_{ls}=
\frac{Q_s}{M_s}(a_{ls}B+b_{ls}D), \quad \dot{h}_{ls}=-\frac{Q_s}{M_s}(a_{ls}H+b_{ls}F).
\label{29_vl}
\end{equation}
One can verify that the Ampere's law without the displacement current is
identically satisfied.

Similarly, from (\ref{12_vl}), (\ref{13_vl}) and (\ref{17_vl})
with (\ref{cb_vl}) in Eq. (\ref{3_vln}), one obtains
\begin{equation} % ------------ (1_w) -------------
\dot{A}x+\dot{B}y+\dot{H} = -x \sum_s Q_s \frac{\dot{a}_{xs}b_{ys}-b_{xs}\dot{a}_{ys}}{\Lambda_s^2} - y \sum_s Q_s
\frac{\dot{b}_{xs}b_{ys}-b_{xs}\dot{b}_{ys}}{\Lambda_s^2}-
\sum_s Q_s \frac{\dot{b}_{xs}h_{ys}-b_{ys}\dot{h}_{xs}}{\Lambda_s^2}
\label{1_work}
\end{equation}
and
\begin{equation} % ------------ (2_w) -------------
\dot{B}x + \dot{D}y +\dot{F} = -x \sum_s Q_s
\frac{\dot{a}_{ys}a_{xs}-a_{ys}\dot{a}_{xs}}{\Lambda_s^2} - y
\sum_s Q_s
\frac{a_{xs}\dot{b}_{ys}-a_{ys}\dot{b}_{xs}}{\Lambda_s^2} -\sum_s
Q_s \frac{\dot{a}_{ys}h_{xs}-a_{xs}\dot{h}_{ys}}{\Lambda_s^2}.
\label{2_work}
\end{equation}
Equating the terms in Eqs. (\ref{1_work}) and (\ref{2_work}) of
similar spatial dependence, we get
\begin{equation} % ------------ (31_vl) -------------
\dot{A}=\sum_sQ_s\Lambda_s^{-2}
(b_{xs}\dot{a}_{ys}-\dot{a}_{xs}b_{ys}), \label{31_vl}
\end{equation}
\begin{equation} % ------------ (32_vl) -------------
\dot{B}=\sum_sQ_s\Lambda_s^{-2}(b_{xs}\dot{b}_{ys} -
\dot{b}_{xs}b_{ys}), \label{32_vl}
\end{equation}
\begin{equation} % ------------ (34_vl) -------------
\dot{D}=\sum_sQ_s\Lambda_s^{-2}(a_{ys}\dot{b}_{xs}-
a_{xs}\dot{b}_{ys}), \label{34_vl}
\end{equation}
\begin{equation} % ------------ (H_vl) -------------
\dot{H}= \sum_s Q_s
\frac{b_{ys}\dot{h}_{xs}-\dot{b}_{xs}h_{ys}}{\Lambda_s^2},
\label{H_vl}
\end{equation}
\begin{equation} % ------------ (F_vl) -------------
\dot{F}= \sum_s Q_s
\frac{a_{xs}\dot{h}_{ys}-\dot{a}_{ys}h_{xs}}{\Lambda_s^2},
\label{F_vl}
\end{equation}
and from (\ref{cb_vl}) the condition
\begin{equation} % ------------ (add_vl) -------------
\sum_sQ_s\Lambda_s^{-2}( b_{xs}\dot{b}_{ys}-\dot{b}_{xs}b_{ys})=
\sum_sQ_s\Lambda_s^{-2}(a_{ys}\dot{a}_{xs}-\dot{a}_{ys}a_{xs}).
\label{add_vl}
\end{equation}

From the mathematical point of view, the equations
(\ref{29_vl})-(\ref{add_vl}) form a closed set that depends on
the parameters $A$, $B$ and $D$. We now consider the physical
meanings of these parameters and the relation (\ref{add_vl}).
Accordingly, we first evaluate
$$\Omega_s = \nabla \times {\bf j}_s.$$
Inserting (\ref{12_vl}) and (\ref{13_vl}), we get
\begin{equation} % ------------ (1_vrt) -------------
\Omega_s = \frac{b_{xs}\dot{b}_{ys} +\dot{a}_{ys}a_{xs}-
\dot{b}_{xs}b_{ys}-a_{ys}\dot{a}_{xs}}{\Lambda_s^2}. \label{1_vrt}
\end{equation}
It follows that the Eq. (\ref{add_vl}) corresponds to the
condition for vortex-free motion. The functions $B(t)$ and $C(t)$
are then related to the vortex component of the electrical field
in (\ref{17_vl}) by $C-B =\sum Q_s\Omega_s$.

On the other hand, combining Eqs. (\ref{31_vl}) and (\ref{34_vl})
and integrating with respect to time, we get
\begin{equation} % ------------ (2_vrt) -------------
A+D =\sum Q_s n_s, \label{2_vrt}
\end{equation}
which shows that $A(t)$ and $D(t)$ are related to the action of
the electrostatic field.

Finally, we should define the moving boundaries $X_s(t)$ and
$Y_s(t)$ for the expanding electron and ion fluids by requiring
that the total number
$$N_s=\int_0^{X_s}\int_0^{Y_s}n_s dxdy$$
of each species of particles is constant since there is no loss or
source of particles in the evolving plasma volume $\Gamma(t)$. For
spatially homogeneous plasma, we have $N_s= X_s Y_s n_s$.

For convenience, we set $N_e(t=0)=N_i(t=0)=1$ in the initial
volume $\Gamma(t=0)$. From the particle conservation condition
$d_t N_s=0$ we obtain
$$Y_s n_s\left[\dot{X_s}+ \frac{\dot{n}_s}{2 n_s}X_s\right]
+ X_s n_s\left[\dot{Y_s}+\frac{\dot{n}_s}{2 n_s}Y_s\right] =0.$$
Since $X_s(t)$ and $Y_s(t)$ are independent, we can set
$$\dot{X_s}+\frac{\dot{n}_s}{2 n_s}X_s=0, \quad \dot{Y_s}+\frac{\dot{n}_s}{2 n_s}Y_s=0$$
and find
\begin{equation} % ------------ (1_XY) -------------
X_s(t)= Y_s(t)= n_s^{-1/2},
\label{1_XY}
\end{equation}
which in view of (\ref{11_vl}) becomes
\begin{equation} % ------------ (XY_vln) -------------
X_s(t)= Y_s(t)=\Lambda_s^{1/2}.
\label{XY_vln}
\end{equation}

\section{Existence of solutions}
We now show that there indeed exist nontrivial solutions of the Eqs. (\ref{29_vl}) -- (\ref{add_vl}). Let us consider the invariants with the coefficients
\begin{equation} % ------------ (nc_vl) -------------
b_{xs}=a_{ys},\quad b_{ys}=a_{xs}, \quad h_{xs}= h_{ys}=0,
\label{nc_vl}
\end{equation}
so that (\ref{add_vl}) is satisfied and Eq. (\ref{32_vl}) becomes
\begin{equation} % ------------ (32_vln) -------------
\dot{B}=\sum_s\frac{Q_s}{a_{xs}^2 - a_{ys}^2}\left[\frac{\dot{a}_{xs}
+ \dot{a}_{ys}}{a_{xs} + a_{ys}} - \frac{\dot{a}_{xs} - \dot{a}_{ys}}{a_{xs} - a_{ys}}\right].
\label{32_vln}
\end{equation}
Eqs. (\ref{31_vl}) and (\ref{34_vl}) become identical:
\begin{equation} % ------------ (AD_vl) -------------
\dot{A}=\dot{D}=\sum_s\frac{Q_s}{a_{xs}^2 - a_{ys}^2}\left[\frac{\dot{a}_{xs}
+ \dot{a}_{ys}}{a_{xs} + a_{ys}} + \frac{\dot{a}_{xs} - \dot{a}_{ys}}{a_{xs} - a_{ys}}\right],
\label{AD_vl}
\end{equation}
so that $A(t)=D(t)$ if $A(t=0)=D(t=0)$. As a result, Eqs. (\ref{29_vl}) reduce to
\begin{equation} % ------------ (29a_vl) -------------
\ddot{a}_{ls}=\frac{Q_s}{M_s} (A a_{xs}+B a_{ys}), \quad
\ddot{a}_{ys}=\frac{Q_s}{M_s}(A a_{ys}+ B a_{xs}).
\label{29a_vl}
\end{equation}
Equation (\ref{34_vl}) can be integrated to
\begin{equation} % ------------ (1_ad) -------------
A=D= \frac{1}{2} \sum_s \frac{Q_s}{ a^2_{ys} - a^2_{xs}}\equiv\frac{1}{2} \sum_s Q_sn_s \ne 0.
\label{1_ad}
\end{equation}

As mentioned, the functions $A(t)$ and $D(t)$ are associated with
the action of electrostatic field ${\bf E}$. However, for
(\ref{nc_vl}) there are no nontrivial quasineutral states, i.e.
the plasma layer always remains charged. To verify this we start
by assuming the opposite, namely $n_e=n_i$. In view of
(\ref{nc_vl}), we can then write
\begin{equation} % ------------ (2_ad) -------------
a^2_{ye} - a^2_{xe}\equiv a^2_{yi} - a^2_{xi},
\label{2_ad}
\end{equation}
and set $A\equiv 0$ in (\ref{29a_vl}). From the reduced equations
(\ref{29a_vl}) one gets
\begin{equation} % ------------ (3_ad) -------------
\frac{M _s}{Q_s}(a_{ys}\ddot{a}_{xs}- a_{xs}\ddot{a}_{ys})=
(a_{ys}^2 - a_{xs}^2).
\label{3_ad}
\end{equation}
The condition (\ref{2_ad}) requires that the right-hand of
(\ref{3_ad}) does not depend on $s$, so that we can set
\begin{equation} % ------------ (4_ad) -------------
a_{ls}= \sqrt{\frac{Q _s}{M_s}}\alpha_l(t),
\label{4_ad}
\end{equation}
where $\alpha_l(t)$ is a function of $t$. However, this form of
$a_{ls}$ cannot satisfy the quasi-neutrality condition
(\ref{2_ad}). Accordingly, the choice (\ref{nc_vl}) cannot
describe quasi-neutral expansion of the plasma slab.

Thus, the ODEs (\ref{32_vln}) -- (\ref{29a_vl}), together with the
initial conditions on the distribution functions, fully determine
the evolution of the plasma, which remains non-neutral for all
$t$. Given the initial values of $a_{ls}$ and $b_{ls}$, one can
numerically integrate these ODEs. The evolution of the
distribution functions is then determined when the explicit forms
of the initial distribution functions
$f_s(I_{xs}(t=0),I_{ys}(t=0))$ are specified.

\section{The reduced case $K=1$}\label{example}
We now consider the degenerate case, where the rank of the matrix
of the algebraic equations (\ref{7_vl}) is unity, or when the
equations are linearly dependent. For simplicity, we shall
concentrate on the case where the distribution function depends
only on one invariant, say $I_s$, or
\begin{equation} % ------------ (35_vl) -------------
f_s=f_s(I_s). \label{35_vl}
\end{equation}
A simple but physically relevant exact solution can be obtained if
we also set $b_s=\lambda a_s$, where $\lambda$ is an arbitrary
constant. Then $I_s$ becomes
\begin{equation} % ------------ (1_rc) -------------
I_s=a_sv_{x}+ \lambda a_sv_{y}-\dot{a}_sx-\lambda\dot{a}_sy + h_s,
\label{1_rc}
\end{equation}
where we have omitted the subscript $l$ (i.e., $a_s=a_{ls}$,
$b_s=b_{ls}$ and $h_s=h_{ls}$). We note that the problem remains
exact and 2D, even though we have used only one invariant and a
specific choice of parameters.

From (\ref{2_cvl}) and (\ref{3_cvl}) one can get the particle
densities and fluxes (see Appendix \ref{AppendixC})
\begin{equation} % ------------ (36_vl) -------------
n_s=\frac{1}{\lambda a_s^2},\quad j_{xs}=\frac{\dot{a}_s x -
h_s}{\lambda a_s^3}, \quad j_{ys}=\frac{\dot{a}_s}{\lambda a_s^3}y.
\label{36_vl}
\end{equation}
The relations (\ref{1_work}) and (\ref{2_work}) then become
\begin{equation} % ------------ (1_wd) -------------
\dot{A}x + \dot{B}y+\dot{H}= - \sum_s Q_s \frac{\dot{a}_s x - h_s}{\lambda a_s^3},
\label{1_wd}
\end{equation}
\begin{equation} % ------------ (2_wd) -------------
\dot{B}x + \dot{D}y + \dot{F}= - \sum_s Q_s \frac{\dot{a}_s y }{\lambda a_s^3}.
\label{2_wd}
\end{equation}
It follows that
\begin{equation} % ------------ (3_wd) -------------
\dot{B} =\dot{F}=0, \quad \dot{A} = \dot{D} =-\sum_s
Q_s\frac{\dot{a}_s}{\lambda a_s^3}, \quad \dot{H} = \sum_s Q_s
\frac{\dot{h}_s}{\lambda a_s^3}. \label{3_wd}
\end{equation}
Integrating the first two relations in (\ref{3_wd}) with respect
to $t$, we find
\begin{equation} % ------------ (4_wd) -------------
B = F= 0, \quad A = D = \frac{1}{2\lambda} \sum_s \frac{Q_s}{ a_s^2},
\label{4_wd}
\end{equation}
so that Eqs. (\ref{29_vl}) become
\begin{equation} % ------------ (44_vl) -------------
\ddot{a}_e=\left(\frac{1}{a_e^2}-\frac{1}{a_i^2}\right)
\frac{a_e}{\lambda},  \quad  \quad \ddot{a}_i=-\delta
\left(\frac{1}{a_e^2}-\frac{1}{a_i^2}\right)\frac{a_i}{\lambda},
\label{44_vl}
\end{equation}
and
\begin{equation} % ------------ (h_vl) -------------
\dot{h}_s=-\frac{Q_s a_s}{M_s}H,
\label{h_vl}
\end{equation}
where $\delta=m_e/m_i$. Substituting (\ref{h_vl}) into the third
equation of (\ref{3_wd}) we obtain
\begin{equation} % ------------ (h2_vl) -------------
\dot{H} = -\sum_s \frac{1}{\lambda M_s a_s^2}H\label{h2_vl}
\end{equation}
which can be integrated to
\begin{equation} % ------------ (h2_vl) -------------
H =
H_0\exp\left[-\frac{1}{\lambda}\int_0^t\left(\frac{1}{a_e^2}
+\frac{\delta}{a_i^2}\right)
dt^{\prime}\right], \label{h3_vl}
\end{equation}
where $H_0$ is an arbitrary constant to be determined by the
initial conditions. Finally, combining (\ref{36_vl}) and
(\ref{1_XY}), we get
\begin{equation} % ------------ (XY_wd) -------------
X_s(t)= Y_s(t)=a_s \lambda^{-1/2},
\label{XY_wd}
\end{equation}
where $\lambda$ is determined by the initial value $N_s(t=0)$. The
evolution of the distribution functions are thereby fully
determined by their initial values $f_s(I_s(t=0))$, where the
invariants $I_s$ are given by the solutions of (\ref{29_vl}). We
note that the coefficients $a_e$ and $a_i$ are the functions
describing the moving boundaries of the electron and ion fluids.
That is, Eqs. (\ref{44_vl}) are the equations of motion for the
corresponding fronts.

\section{The behavior at short and long times}
The ODEs (\ref{44_vl}) can be solved numerically when $\delta$ and
$\lambda$, as well as $a_e$ and $a_i$ and their time derivatives
at $t=0$ are given, so that the solutions depend only on these
initial conditions. Typical solutions are shown in Fig.\ref{f1}:
(a) free expansion of the plasma slab, (b) expansion with
large-amplitude oscillations, and (c) contracting plasma slab with
oscillating electron front. Numerical investigation also allows us
to obtain an empirical relation
\begin{equation} % ------------ (chp_vl) -------------
D_- a_i < a_e < D_+ a_i,
\label{chp_vl}
\end{equation}
where $D_->0$ and $D_+>0$ are constants. This relation reflects
the electrostatic interaction between the ion and electron fluids.

\renewcommand{\caption}{{\bf Figure \arabic{graf}. }}
\begin{figure}   %[htbp]
%\vspace{16.5pc}
\begin{center}
\includegraphics[width=10cm]{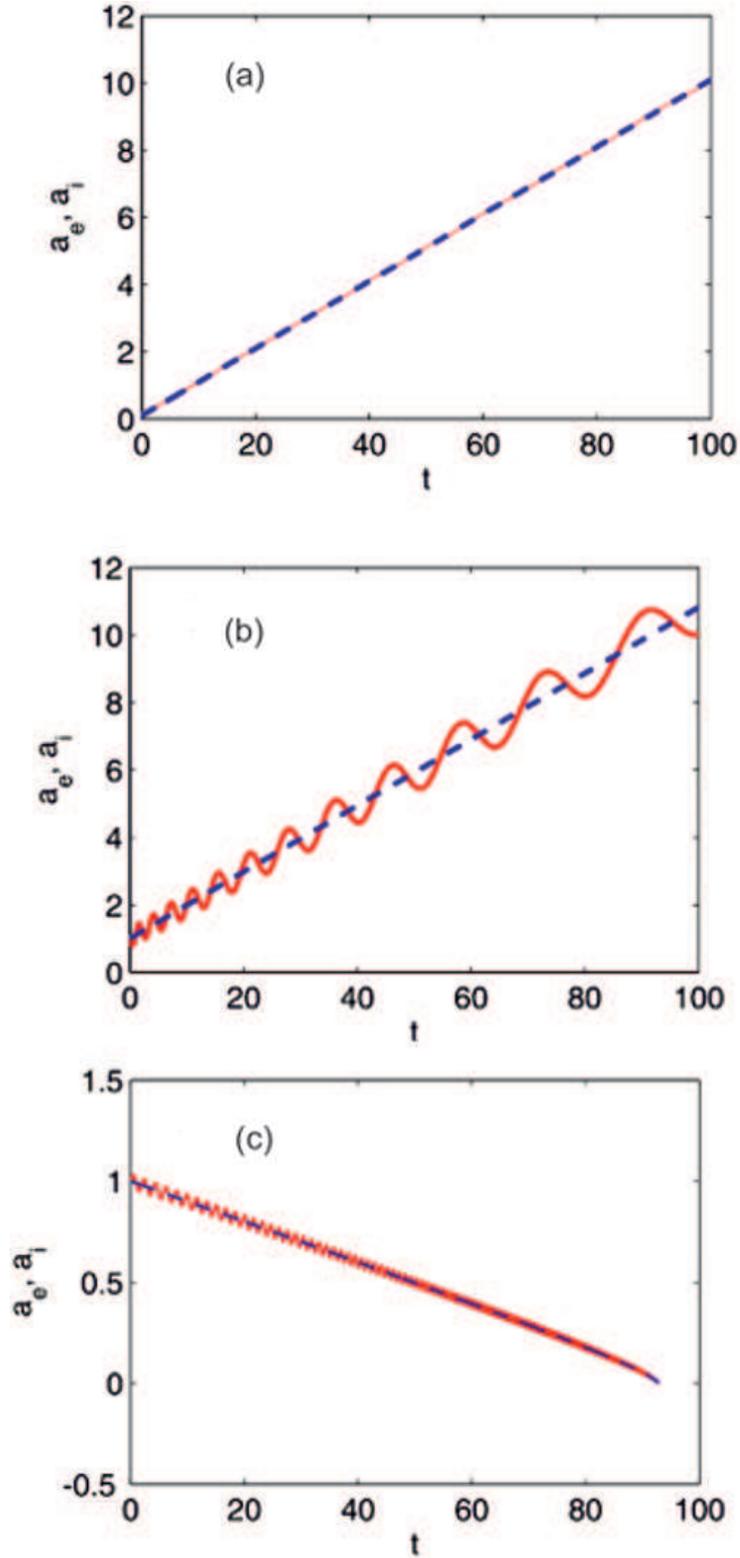}
\refstepcounter{graf}\label{f1}
\end{center}
\caption{(Color online.) Evolution of the ion and electron fronts $a_i(t)$
(blue dotted curve) and $a_e(t)$ (red solid curve) respectively for the
different initial data: (a) - $a_i(0)=a_e(0)=0.1$,
$\dot{a}_e(0)=10^{-6}$, $\dot{a}_i(0)=0.1$; (b) - $a_i(0)=a_e(0)=1$,
$\dot{a}_e(0)=-0.7$, $\dot{a}_i(0)=0.1$; (c) - $a_i(0)=a_e(0)=1$,
$\dot{a}_e(0)=0.1$, $\dot{a}_i(0)=-0.01.$}
\end{figure}
One can give a qualitative analysis of the expansion dynamics at
large times. Combining the Eqs. (\ref{44_vl}) we get
$$\frac{\ddot{a_e}}{a_e}+ \frac{1}{\delta}\frac{\ddot{a_i}}{a_i}=0\/,$$
which after integration with the initial conditions (\ref{2_vlj})
yields
\begin{equation} % ------------ (18_coup) -------------
{\dot{a}_e \over a_e}+ {1\over \delta}{\dot{a}_i \over a_i}+
\int_0^t \left[\left({\dot{a_e} \over a_e}\right)^2+ {1\over
\delta}\left({\dot{a}_i \over a_i}\right)^2\right]
dt^{\prime}=0\/. \label{18_coup}
\end{equation}
Further integration gives
\begin{equation} % ------------ (19_coup) -------------
a_e {a}_i^{1/\delta}=\exp\left[-\int_0^t \theta(a_e, a_i) dt^{\prime}\right]\/,
\label{19_coup}
\end{equation}
where
$$\theta(a_e, a_i)= \int_0^t \left[\left({\dot{a_e}
\over a_e}\right)^2+ {1\over \delta}\left({\dot{a}_i \over
a_i}\right)^2\right] dt^{\prime} > 0.$$ Taking into account
(\ref{chp_vl}), we can rewrite (\ref{19_coup}) as
\begin{equation} % ------------ (20_coup) -------------
a_i < D_-^{-\frac{\delta}{1 +\delta}}
\exp\left[-\frac{\delta}{1 +\delta}
\int_0^t \theta(a_e, a_i) dt^{\prime}\right]\/.
\label{20_coup}
\end{equation}
This relation implies that $a_e$ and $a_i$, and thus $X_s$ and
$Y_s$, are always bounded. In view of the Chaplygin comparison
theorems [\cite{chaplygin}], similar behavior for the more general
case $K=2$ can be expected.

\section{Discussion and conclusion}
In contrast to the asymptotic stationary solution, namely the
Maxwell distribution, of the Boltzmann and other equations
including collision or velocity-space diffusion effects, the
Vlasov equation can have an infinite number of asymptotic states
[\cite{clemmow69}], depending on the initial distribution. For
example, Refs. [\cite{dz90}--\cite{m97}] showed that even though
small-amplitude electric fields can be damped and eventually
vanish, sufficiently large-amplitude perturbations can evolve into
wave-like or other states. The results here belong to the latter
class. The Vlasov system possesses such a property because it
precludes direct particle-particle collisions that tend to
randomize the particle velocities, and the interaction via the
self-consistent electrostatic field cannot change the system
entropy. However, one can still compare the macroscopic quantities
(velocity-space moments of the distribution function) and the
electrostatic field with that obtained from the corresponding
fluid models. In fact, some of our results on the evolution of
initially confined plasmas are similar to phenomena predicted by
the latter [\cite{ksy_1}-\cite{wywk16}].

In this paper we have considered the properties and expansion of a
collisionless plasma slab with anisotropic nonequilibrium particle
distributions. We obtained fully nonlinear time-dependent, 2D
solutions of the Vlasov-Maxwell equations by invoking the Jeans'
theorem. In contrast to most existing works invoking the latter,
here the invariants of motion used to construct the distribution
function are linear combinations of the phase-space variables, but
the coefficients are time-dependent and governed ODEs. That is,
they are not related to the traditional conservation laws such as
that for energy and momentum. The plasma density, flux, as well as
the electrostatic field then depend on the form of the invariants
as well as how they appear in the distribution function, as can be
seen from the relations (\ref{17_vl}), (\ref{11_vl}),
(\ref{12_vl}) and (\ref{13_vl}). The solutions then describe
nonequilibrium plasma flows, where imbalance among the different
degrees of freedom leads to a preferred direction of evolution in
the phase space.

We emphasize that the solutions, including the highly simplified
but physically nontrivial case $K=1$, are mathematically exact and
are also valid for open systems, including that with
nonconservative space and time dependent external forces. One can
expect that similar results can also be found for higher
dimensional systems. Finally, we note that by using polynomial
(instead of linear) forms of the invariants, the dynamics of other
systems of physical interest [\cite{arg05,ll,struc}] can also be
considered, such as that of a vortex system [\cite{ES06,S92}].

\begin{acknowledgments}
The present paper stems from the questions put by a Journal of Plasma Physics referee on [\cite{pecsili_16}]. The authors would like to express their profound gratitude to the referee for the valuable remarks and suggestions on the lines of further researches. M.Y.Y. was supported by the National Natural Science Foundation of China (11374262, 11475147) and the State Key Laboratory of High Field Laser Physics at SIOM.
\end{acknowledgments}

\appendix
\section{Particle densities and fluxes for $K=2$}\label{AppendixB}
Here we show how the time-dependent coefficients of the invariants
$I_{xs}$ and $I_{ys}$ in (\ref{7_vl}) are related to the particle
densities and fluxes of the initially bounded plasma. From
(\ref{7_vl}), we have
\begin{eqnarray} % ------------ (1_avl) -------------
&\Lambda_sv_x(I_{xs},I_{xs})=b_{ys}I_{xs}-b_{xs}I_{ys}+\nonumber \\
&(b_{ys}\dot{a}_{xs}-b_{xs}\dot{a}_{ys})x +(b_{ys}\dot{b}_{xs}-
b_{xs}\dot{b}_{ys})y+ b_{xs}h_{ys}-b_{ys}h_{xs}, \label{1_avl}
\end{eqnarray}
\begin{eqnarray} % ------------ (2_avl) -------------
&\Lambda_sv_y(I_{xs},I_{xs})= a_{xs}I_{ys}-a_{ys}I_{xs}+\nonumber\\
&(a_{xs}\dot{a}_{ys}-a_{ys}\dot{a}_{xs})x +(a_{xs}\dot{b}_{ys}-
a_{ys}\dot{b}_{xs})y + a_{ys}h_{xs}-a_{xs}h_{ys}, \label{2_avl}
\end{eqnarray}
where $\Lambda_s=a_{xs}b_{ys}-a_{ys}b_{xs}$.

%Using the transformation $dv_x dv_y={\cal J}dI_{xs}dI_{ys}$,
In terms of the invariants, we can express the density as
\begin{eqnarray} % ------------ (4_avl) -------------
n_s&=\int f_s(I_{xs},I_{ys})dv_xdv_y\label{4_avl}\\
&=\Lambda_s^{-1}\int f_s(I_{xs},I_{ys})dI_{xs}dI_{ys}. \nonumber
%%n_s&=\int f_s(I_{xs},I_{ys}){\cal J}dI_{xs}dI_{ys}\label{4_avl}\\
%%&=\Lambda_s^{-1}\int f_s(I_{xs},I_{ys})dI_{xs}dI_{ys}. \nonumber
\end{eqnarray}
where we have used the transformation Jacobian ${\cal
J}={D(v_x,v_y)/D(I_{xs},I_{ys})}={1/\Lambda_s}$. In view of the
initial or normalization condition, one can see that the plasma
density is $n_s={\Lambda_s^{-1}}$. That is, the plasma indeed
remains homogeneous during its evolution.

Similarly, for the macroscopic flux we have
\begin{eqnarray} % ------------ (6_avl) -------------
&j_{xs}=\int v_x(I_{xs},I_{xs})f_s(I_{xs},I_{ys})\Lambda_s^{-1}
dI_{xs}dI_{ys} \label{6_avl}\\
&=\Lambda_s^{-2}\left[(\dot{a}_{xs}b_{ys}-b_{xs}\dot{a}_{ys})x+(\dot{b}_{xs}
b_{ys}-b_{xs}\dot{b}_{ys})y + b_{xs}h_{ys}-b_{ys}h_{xs}\right],\nonumber
\end{eqnarray}
and
\begin{eqnarray} % ------------ (7_avl) -------------
&j_{ys}=\int v_y(I_{xs},I_{xs})f_s(I_{xs},I_{ys})\Lambda_s^{-1}
dI_{xs}dI_{ys} \label{7_avl}\\
&=\Lambda_s^{-2}[(\dot{a}_{ys}a_{xs}-a_{ys}\dot{a}_{xs})x
+(a_{xs}\dot{b}_{ys}-a_{ys}\dot{b}_{xs})y + a_{ys}h_{xs}-a_{xs}h_{ys}],\nonumber
\end{eqnarray}
where we have again used the initial condition. Thus, the
macroscopic flow parameters are rather complicated functions of
the structure coefficients appearing in (\ref{7_vl}).

\section{Particle densities and fluxes for $K=1$} \label{AppendixC}
Here we obtain the macroscopic densities and fluxes for $K=1$ by
using the solution (\ref{35_vl}) with (\ref{1_rc}). Accordingly,
we have
$$n_s=\int f_s(a_sv_{x}+\lambda a_sv_{y}-\dot{a}_sx-\lambda \dot{a}_s y +h_s)dv_x dv_y=
\frac{1}{\lambda a_s^2} \int f_s(\xi + \eta)d\xi d\eta,$$ where we
have used
$$\xi= a_sv_{x}-\dot{a}_s x + h_s, \quad\quad\eta=\lambda(a_sv_{y}-\dot{a}_s y).$$
In view of the initial condition (\ref{2_vlj}) we obtain
\begin{equation} % ------------ (2_cvl) -------------
n_s=\frac{1}{\lambda a_s^2}. \label{2_cvl}
\end{equation}
The corresponding macroscopic fluxes are
$$j_{x s}= \int v_x f_s(a_sv_{x}+ \lambda a_sv_{y}-\dot{a}_sx-\lambda \dot{a}_s y
+ h_s)dv_x dv_y$$
and
$$j_{y s}= \int v_y f_s(a_sv_{x}+ \lambda a_sv_{y}-\dot{a}_sx-\lambda
\dot{a}_s y +h_s)dv_x dv_y.$$
These can be rewritten as
$$j_{x s}= \frac{1}{\lambda a_s^3}\left(\int \xi f(\xi+\eta)d\xi d\eta
+(\dot{a}_s x - h_s)
\int f(\xi+\eta)d\xi d\eta \right),$$
$$j_{y s}=\frac{1}{\lambda^2 a_s^3}\left(\int \eta f(\xi+\eta)d\xi d\eta+
\lambda \dot{a}_s y \int f(\xi+\eta)d\xi d\eta\right).$$
Applying the initial conditions (\ref{2_vlj}), we find
\begin{equation} % ------------ (3_cvl) -------------
j_{x s}= \frac{\dot{a}_s - h_s }{\lambda a_s^3 }x, \quad j_{y
s}=\frac{\dot{a}_s}{\lambda a_s^3}y. \label{3_cvl}
\end{equation}
We note that for the distribution (\ref{35_vl}) with (\ref{1_rc}),
the relation (\ref{2_cvl}) is not unique. It is the simplest
nontrivial choice. One can obtain other results for $j_{x s}$ and
$j_{y s}$ if different $\xi$ and $\eta$ are used.

\end{document}